\documentclass[prb,preprintnumbers,superscriptaddress,amsmath,amssymb,floatfix,aps,twocolumn]{revtex4}
\usepackage{placeins}

\usepackage{graphicx}

\usepackage{lipsum} 
\usepackage{comment}
\usepackage{lineno}

\usepackage{xcolor}

\usepackage{ulem}

\usepackage{afterpage}

\usepackage[version=4]{mhchem}

\usepackage{amsmath}
\usepackage{amsbsy}
\usepackage{float}
\usepackage[utf8x]{inputenc}

\begin{document}

\title{Strain-Induced Half-Metallicity and Giant Wiedemann-Franz Violation in Monolayer NiI$_2$}

\date{\today} 

\author{J. W. Gonz\'alez}
\email{jhon.gonzalez@uatonf.cl}
\affiliation{Departamento de Física, Universidad de Antofagasta, Av. Angamos 601, Casilla 170, Antofagasta, Chile}

\author{L. Rosales}
\email{luis.rosalesa@usm.cl}
\affiliation{Departamento de Física, Universidad Técnica Federico Santa María, Av. España 1680, Casilla 110V, Valparaíso, Chile}

\begin{abstract}
Reversible control of spin-dependent thermoelectricity via mechanical strain provides a platform for next-generation energy harvesting and thermal logic circuits. Using first-principles and Boltzmann transport calculations, we demonstrate that monolayer NiI$_2$ undergoes a strain-driven semiconductor-to-half-metal transition, enabled by the selective closure of its spin-down band gap while preserving a robust ferromagnetic ground state. Remarkably, this transition is accompanied by a giant, non-monotonic violation of the Wiedemann-Franz law, with the Lorenz number enhanced up to $7.17\,L_0$. This anomaly arises from a strain-sensitive hybridization between Ni-$d$ and I-$p$ orbitals, leading to spin-polarized transport channels and decoupling of heat and charge currents. These properties make NiI$_2$ a promising candidate for mechanically gated spin-caloritronic devices and thermal logic elements, where reversible control of heat and spin flow is essential. Our findings position NiI$_2$ as a model system for exploring non-Fermi-liquid transport and for realizing strain-tunable, energy-efficient functionalities in low-dimensional platforms.
\end{abstract}

\maketitle

\section{\label{sec:intro} Introduction}

Two-dimensional (2D) materials have revolutionized condensed matter physics and materials science due to their extraordinary electronic, magnetic, optical, and thermal properties. These characteristics can be finely tuned through external stimuli such as mechanical strain, electric and magnetic fields, or chemical functionalization~\cite{deilmann2023optical, qi2023recent}. The discovery of intrinsic two-dimensional magnets has significantly broadened the scope for investigating fundamental physical phenomena and for advancing the design of next-generation spintronic and spin-caloritronic devices~\cite{zhang2021two, elahi22review}. Among these emerging materials, transition metal dihalides (TMDHs), such as NiI$_2$, have attracted particular interest due to their intrinsic magnetic properties and distinctive layered van der Waals structure~\cite{jiang2022general}.

Monolayer NiI$_2$ has garnered attention due to its unusual combination of semiconducting behavior, strong spin polarization, and complex magnetic ordering, including helimagnetism and multiferroicity even at the monolayer scale~\cite{wang2024orientation, nii2_multiferroic_osti, nii2_helimagnetic_arxiv}. These properties, coupled with recent advances in synthesis and exfoliation techniques, have enabled the production of high-quality monolayer samples, facilitating detailed experimental and theoretical investigations~\cite{jiang2022general, wang2024orientation}. These attributes make NiI$_2$ an excellent platform for studying the interplay between charge, spin, and heat transport in low-dimensional systems, a central theme in the development of thermoelectric and spin-caloritronic devices. The performance of thermoelectric materials is closely tied to the entropy per charge carrier, and magnetic systems can enhance efficiency by leveraging the spin degree of freedom~\cite{canetta2024impact}. Moreover, the possibility of observing spin-caloritronic effects such as the spin Seebeck effect, where a spin current is generated from a thermal gradient, further underscores the relevance of 2D magnetic materials in energy harvesting and spintronic applications~\cite{spinseebeck_2d_review, crl3_spin_seebeck}.

\begin{figure}[]
\centering
\includegraphics[clip,width=0.48\textwidth,angle=0]{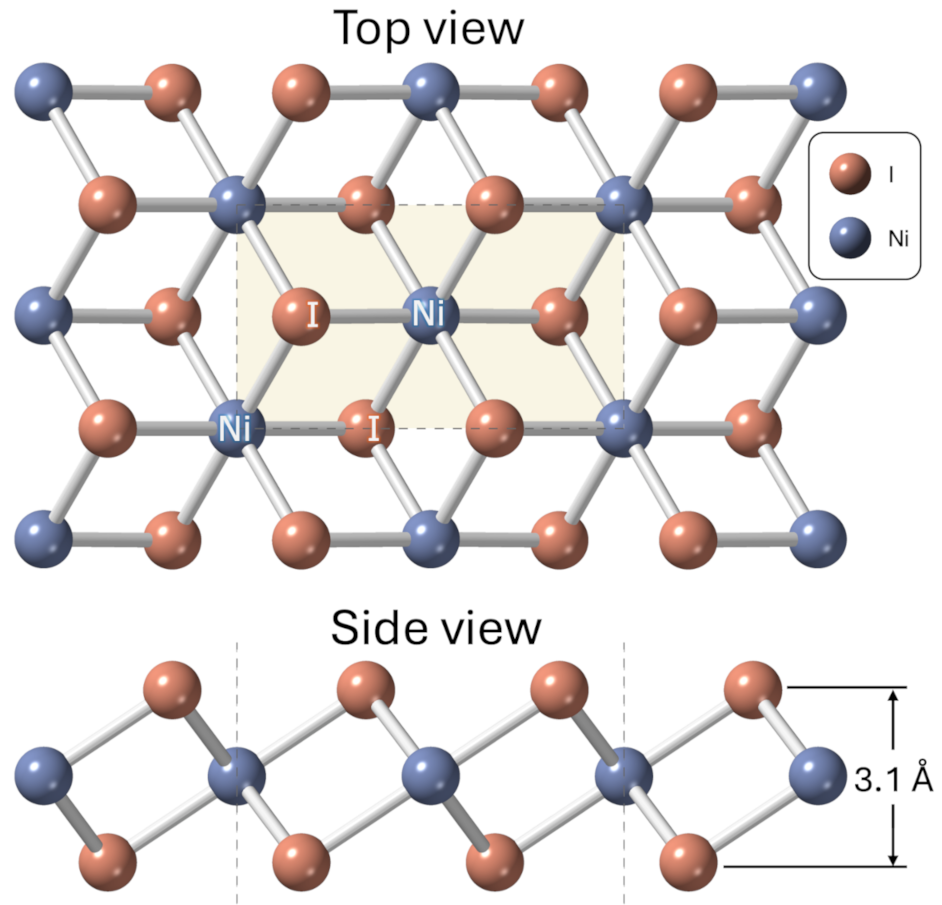} 
\caption{Schematic representation of monolayer NiI\(_2\) in the ferromagnetic configuration. The rectangular supercell has lattice parameters \( a = 6.88\,\text{\AA} \) and \( b = 3.97\,\text{\AA} \), and contains four iodine atoms and two nickel atoms. The nearest-neighbor Ni--Ni distance is 3.9\,\text{\AA}, and the layer thickness, defined by the vertical I--I separation, is 3.1\,\text{\AA}.}
\label{Fig:scheme}
\end{figure}

Strain engineering has emerged as a highly effective and reversible strategy for modulating the physical properties of 2D materials~\cite{qi2023recent}. In layered magnetic semiconductors like NiI$_2$, biaxial strain can modify the stability order between competing magnetic phases, reshape the band structure, and strongly influence carrier mobility and thermoelectric performance~\cite{leon2020strain, gonzalez2021strain, zhang2021two, gonzalez2019highly}. This exceptional tunability is particularly valuable for optimizing thermoelectric figures of merit, as key transport coefficients, such as electrical conductivity and the Seebeck coefficient, are highly sensitive to the electronic structure and scattering processes. Additionally, strain serves as a powerful tool for probing and enhancing spin-caloritronic effects in 2D magnets~\cite{zhang2015strain, elahi22review, canetta2024impact}.

In this work, we perform comprehensive density functional theory (DFT) calculations combined with semiclassical Boltzmann transport theory to systematically investigate the electronic, magnetic, and thermoelectric properties of monolayer NiI$_2$ under biaxial strain. We analyze the energetic competition between ferromagnetic (FM) and antiferromagnetic (AF) configurations, trace the evolution of the spin-resolved band structure, and quantify the strain dependence of key thermoelectric coefficients, including the Seebeck coefficient, electrical conductivity, and power factor. Our findings demonstrate that mechanical strain can induce a semiconductor-to-metal transition in the FM phase, enhance spin-polarized transport, and provide a versatile route to control multifunctional behavior in NiI$_2$ monolayers. These results highlight the promise of NiI$_2$ as a candidate for next-generation spintronic and thermoelectric technologies.

\section{Methodology}
We study the physical properties of monolayer NiI$_2$ under biaxial strain using spin-polarized density functional theory (DFT) as implemented in \textsc{Quantum-ESPRESSO} (v.~7.4.1)~\cite{QE,QE2}. The calculations employ the PBE formulation of the GGA~\cite{PBE} and optimized norm-conserving Vanderbilt pseudopotentials with PBEsol from the \textsc{PSlibrary}~\cite{pseudos}. We use a plane-wave cutoff of 680\,eV (50\,Ry) and include a vacuum layer of 18\,\AA{} along the $c$-axis to avoid spurious interlayer interactions.
We apply a Hubbard-like correction within the rotationally invariant Dudarev formalism~\cite{Dudarev}. We adopt $U = 2$\,eV for Ni $3d$ states, a value supported by high-throughput studies~\cite{Pakdel2024}, machine-learning predictions~\cite{Yu2020}, and prior LDA+$U$ calculations on NiI$_2$ that reproduce experimental gaps and thermoelectric properties~\cite{Prayitno2023,An2022}. This choice balances gap opening with carrier delocalization, thereby improving accuracy without the computational cost associated with hybrid functionals.
We perform structural relaxations using a $9\times9\times1$ Monkhorst-Pack $\mathbf{k}$-point grid, increasing to $21\times21\times1$ for static and density of states calculations to ensure convergence. We relax atomic positions and lattice vectors until residual forces fall below $10^{-3}$\,eV/\AA{} and total energy changes are smaller than $10^{-8}$\,eV. A Fermi-Dirac smearing of 0.001\,eV (0.0001\,Ry) aids convergence.

We evaluate thermoelectric transport properties using semiclassical Boltzmann transport theory within the constant relaxation time approximation (RTA), as implemented in \textsc{BoltzTraP2}~\cite{BoltzTraP,BoltzTraP2}. This approach reliably captures strain-induced trends~\cite{gonzalez2019highly,gonzalez2021strain,shamim2025thermoelectric}. We compute the Seebeck coefficient ($S$), electrical conductivity ($\sigma/\tau$), electronic thermal conductivity ($\kappa_e/\tau$), and power factor ($\mathrm{PF}$) as functions of chemical potential and temperature. Assuming a constant relaxation time $\tau$ across energies and strain values allows us to isolate strain-induced variations and spin-dependent trends, as absolute values of $\sigma$ and PF remain scaled by $\tau$.
Finally, we compute the lattice thermal conductivity using \textsc{Phonopy}~\cite{phonopy} and \textsc{Phono3py}~\cite{phono3py}, including anharmonic phonon-phonon interactions via a $2\times2\times1$ supercell. The Wigner transport equation~\cite{simoncelli2022wigner} accounts for both particle-like and wave-like heat transport. Comparison between full scattering solutions and the phonon RTA shows only minor differences, validating the RTA as a reliable approximation for monolayer NiI$_2$.

We compute electronic transport coefficients under the constant relaxation time approximation (RTA), where the key quantity is the transport distribution function $\Sigma(E)$:
\begin{equation}
   \Sigma(E) \;=\; \frac{1}{N_k} \sum_{n,\mathbf{k}} 
   \tau \;\bigl|\mathbf{v}_{n,\mathbf{k}}\bigr|^2\, \delta\bigl(E - \varepsilon_{n,\mathbf{k}}\bigr),
   \label{eq:Sigma}
\end{equation}
with $\tau$ the relaxation time, $N_k$ the number of $\mathbf{k}$-points, $\varepsilon_{n,\mathbf{k}}$ the band energies, and $\mathbf{v}_{n,\mathbf{k}} = \hbar^{-1} \nabla_{\mathbf{k}} \varepsilon_{n,\mathbf{k}}$ the group velocity. The Dirac delta ensures that only states at energy $E$ contribute to $\Sigma(E)$.

Using $\Sigma(E)$, the electrical conductivity $\sigma$, Seebeck coefficient $S$, and power factor $\mathrm{PF}$ are given by:
\begin{equation}
    \sigma(T,\mu) = e^{2} \int \left(- \frac{\partial f}{\partial E}\right) \Sigma(E)\, dE,
    \label{eq:sigma}
\end{equation}
\begin{equation}
    S(T,\mu) = \frac{1}{e\,T} \frac{\int (E - \mu)\,\left(- \frac{\partial f}{\partial E}\right) \Sigma(E)\, dE}{\int \left(- \frac{\partial f}{\partial E}\right) \Sigma(E)\, dE},
    \label{eq:seebeck}
\end{equation}
\begin{equation}
    \mathrm{PF}(T,\mu) = S^2(T,\mu) \;\sigma(T,\mu),
    \label{eq:PF}
\end{equation}
where $e$ is the elementary charge, $\mu$ the chemical potential, $T$ the temperature, and $f(E,\mu,T)$ the Fermi-Dirac distribution.

To evaluate the validity of the Wiedemann-Franz law\citep{chester1961law} under strain, we compute the energy-dependent Lorenz number\cite{usui2017enhanced,tu2023wiedemann}, defined as
\begin{equation}
L(E) = \frac{\kappa_e(E)}{\sigma(E)\, T},
\label{eq:L}
\end{equation}
where $\kappa_e$ is the electronic contribution to the thermal conductivity, $\sigma$ is the electrical conductivity, and $T$ is the absolute temperature. Within the framework of Fermi liquid theory and elastic scattering, $L(E)$ approaches a universal value known as the Sommerfeld constant,
\begin{equation}
L_0 = \frac{\pi^2}{3} \left( \frac{k_B}{e} \right)^2 \approx 2.44 \times 10^{-8}\, \mathrm{W}\cdot\Omega/\mathrm{K}^2,
\end{equation}
where $k_B$ is the Boltzmann constant and $e$ is the elementary charge. Deviations of $L(E)$ from $L_0$ indicate a breakdown of the Wiedemann-Franz relation, which may arise from inelastic scattering, spin-dependent transport, or strong band structure effects\cite{usui2017enhanced,tu2023wiedemann}.

In our transport calculations, we employ the constant relaxation time approximation (RTA), a widely used method in semiclassical Boltzmann theory. This approach allows efficient estimation of electrical conductivity (\( \sigma/\tau \)) and Seebeck coefficient (\( S \)) based on the electronic structure, assuming a single average and isotropic relaxation time \( \tau \). While \( \tau \) may in principle depend on energy, temperature, or external strain, such variations typically affect absolute values but not the qualitative trends we aim to capture.
Our study focuses on the relative evolution of transport properties under strain, for which the RTA remains an appropriate and practical framework. More accurate predictions would require explicit calculations of strain-dependent scattering rates (e.g., from electron–phonon coupling), which lie beyond the scope of this work.

\begin{figure}[htbp!]
\centering
\includegraphics[clip,width=0.45\textwidth,angle=0]{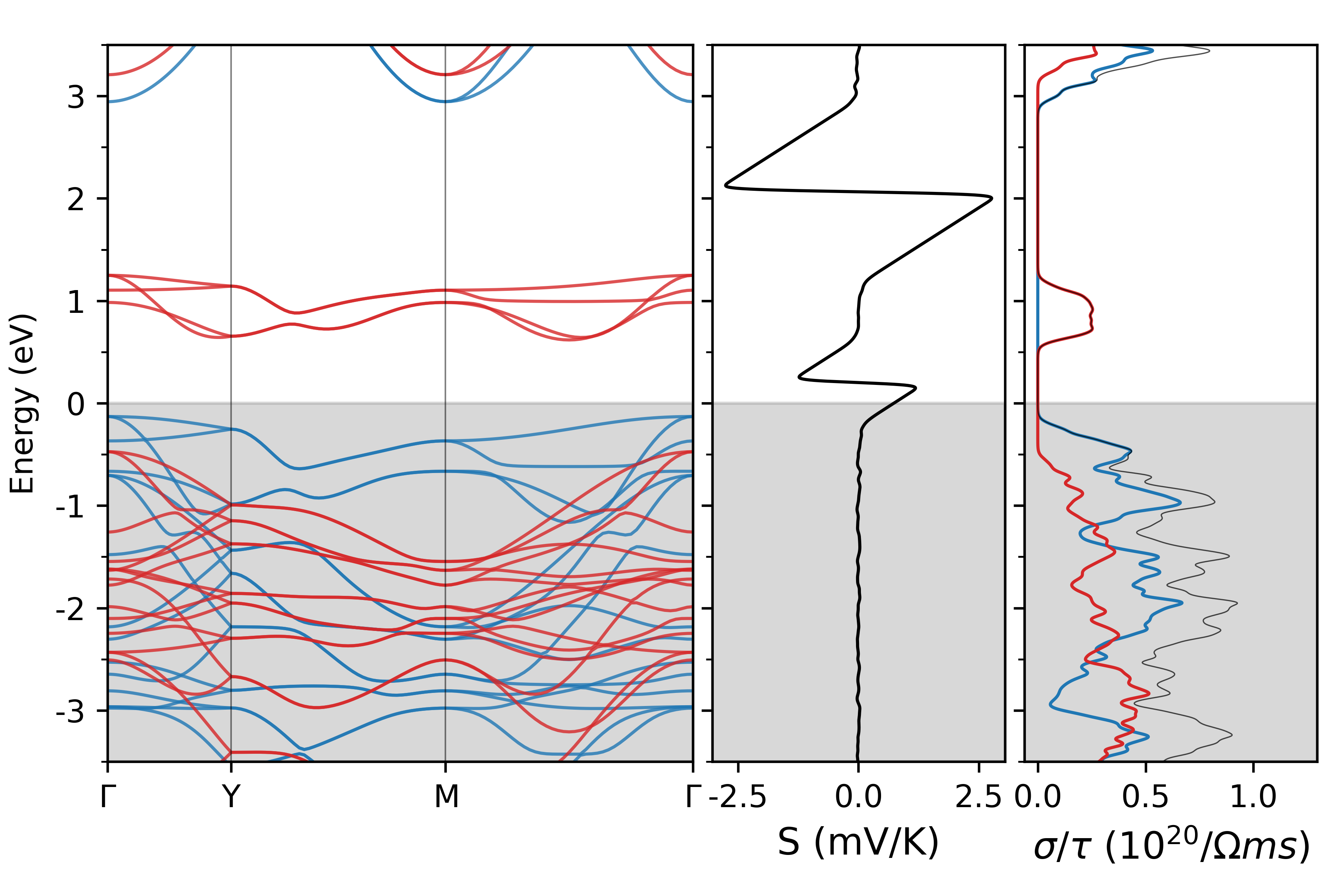} 
\includegraphics[clip,width=0.45\textwidth,angle=0]{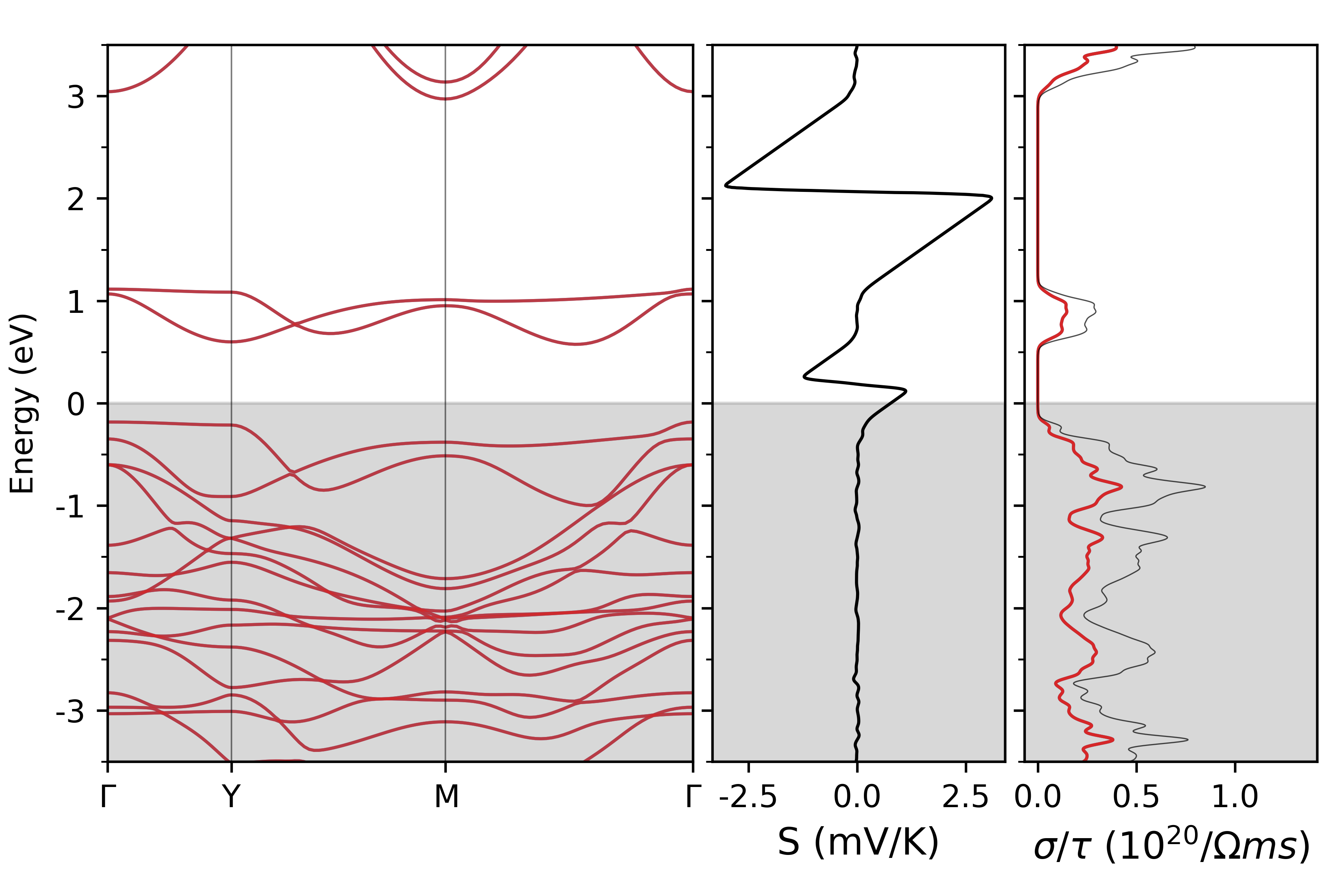} 
\caption{Electronic band structures (left panels), Seebeck coefficients \( S \) (middle panels), and electrical conductivities \( \sigma/\tau \) (right panels) for the ferromagnetic (\textbf{top row}) and antiferromagnetic (\textbf{bottom row}) configurations of monolayer NiI\(_2\). Energies are referenced to the Fermi level, set at 0~eV. Shaded regions indicate the occupied states. Blue and red curves represent the spin-up and spin-down channels, respectively. In the right panels, the gray line denotes the total electrical conductivity, while the colored lines correspond to the spin-resolved contributions.}
\label{Fig:bands}
\end{figure}

\section{\label{sec:mag-strain} Magnetism and Strain}

\subsection{Monolayer without  strain}
In Figure~\ref{Fig:scheme} we show a schematic representation of the NiI$_2$ monolayer in its FM configuration. NiI$_2$ crystallizes in the space group $P\overline{3}m1$ (No.~164), a layered structure also found in other transition-metal dihalides, such as MX$_2$ compounds with M = Mn, Ni, and Fe and X = I, Br, and Cl. In our rectangular supercell, the lattice constants are $a = 6.88\,\text{\AA}$ and $b = 3.97\,\text{\AA}$, containing four I atoms and two Ni atoms. A vacuum region of $18\,\text{\AA}$ along the $z$-direction is included to avoid spurious interactions between periodic images. In the FM configuration, the nearest-neighbor Ni--Ni distance is $3.91\,\text{\AA}$, and the monolayer thickness, defined by the vertical I--I separation, is $3.1\,\text{\AA}$. With a Hubbard parameter $U$ of $2\,\text{eV}$, each Ni atom exhibits a magnetic moment of approximately $1.8\,\mu_B$. In contrast, for the AF configuration, the structure retains the same overall symmetry but with slightly different lattice parameters ($a = 6.87\,\text{\AA}$ and $b = 3.97\,\text{\AA}$). Consequently, the Ni--Ni distance in the AF configuration increases to $3.97\,\text{\AA}$, while the layer thickness decreases marginally to $3.06\,\text{\AA}$, with the Ni magnetic moment remaining similar to that in the FM configuration.

We calculate the electronic properties of the NiI$_2$ monolayer without mechanical strain. In Figure~\ref{Fig:bands}, we present the electronic band structures (left panels), the Seebeck coefficients $S$ (middle panels), and the electrical conductivity $\sigma/\tau$ (right panels) for NiI$_2$ in both ferromagnetic (FM, top) and antiferromagnetic (AF, bottom) configurations. All energies are referenced to the Fermi level, set at $0,\text{eV}$, with the shaded regions indicating the occupied bands below the Fermi level.
In both magnetic configurations, the band structure reveals the material's wide bandgap semiconductor behavior, with distinct indirect energy gaps for each spin component. In the AF NiI$_2$ monolayer, the band structure remains spin-degenerate, resulting in a band gap of approximately 0.76 eV. In contrast, in the FM phase, Kramers' theorem no longer enforces spin degeneracy\cite{ramazashvili2009kramers}, so the spin-up and spin-down channels exhibit separate sets of bands. As a result, we observe a spin-up gap of 3.1 eV, a spin-down gap of 1.1 eV, and a total band gap of 0.75 eV, which corresponds to the energy difference between the highest occupied band and the lowest unoccupied band across both spin channels. 

The energy gap behavior around the Fermi level for the FM configuration defines the Seebeck coefficient, which changes sign and magnitude as the chemical potential shifts through the bands edges. In particular, $S(T,\mu)$ (middle panels, Fig.~\ref{Fig:bands}) reaches its highest absolute values near the edges of the valence and conduction bands, where the transport distribution function  $\Sigma(E)$  has the most pronounced variations. The corresponding electrical conductivity $\sigma/\tau$ (right panel) is completely spin polarized and relatively low within the gap region, increasing significantly as the chemical potential aligns with the conduction bands above $E_F$ or the valence bands below $E_F$. We will explore this spin-dependent properties of the FM NiI$_2$ monolayer in more detail further in this paper.

For the AF configuration (bottom panels, Fig.~\ref{Fig:bands}), the overall shape of the band structure remains similar to that of the FM phase, reflecting the same underlying crystal symmetry. However, subtle differences in the band dispersions and the position of the band edges can be observed. 
These shifts lead to variations in both the Seebeck coefficient and the conductivity. In particular, the AF configuration exhibits a slightly larger Ni--Ni distance and a modified layer thickness, which can alter the bandwidths and effective masses of carriers. Such geometric modifications often translate into a modified electronic structure, 
potentially shifting the energy levels and modifying the density of states near $E_F$. As a result, the peak values of $S$ and $\sigma/\tau$ may appear at slightly different energies relative to the Fermi level compared to the FM case.

Given the presence of heavy iodine atoms in monolayer NiI$_2$, we have assessed the role of spin–orbit coupling (SOC) on its electronic structure. To this end, additional band structure calculations were carried out for the ferromagnetic configuration with SOC included. The results show that SOC induces minor band splittings at energies well below the Fermi level (particularly below –1.5 eV), but has a negligible impact on both the bandgap and the states near the Fermi level. Importantly, the magnetization was set along the out-of-plane (z) direction, consistent with the easy axis identified in previous studies\cite{han2020enhanced}, which report a strong perpendicular magnetic anisotropy in monolayer NiI$_2$. Consequently, neither the magnetic ground state nor the thermoelectric properties are significantly affected.

\begin{figure*}[]
\centering
\includegraphics[clip,width=0.98\textwidth,angle=0]{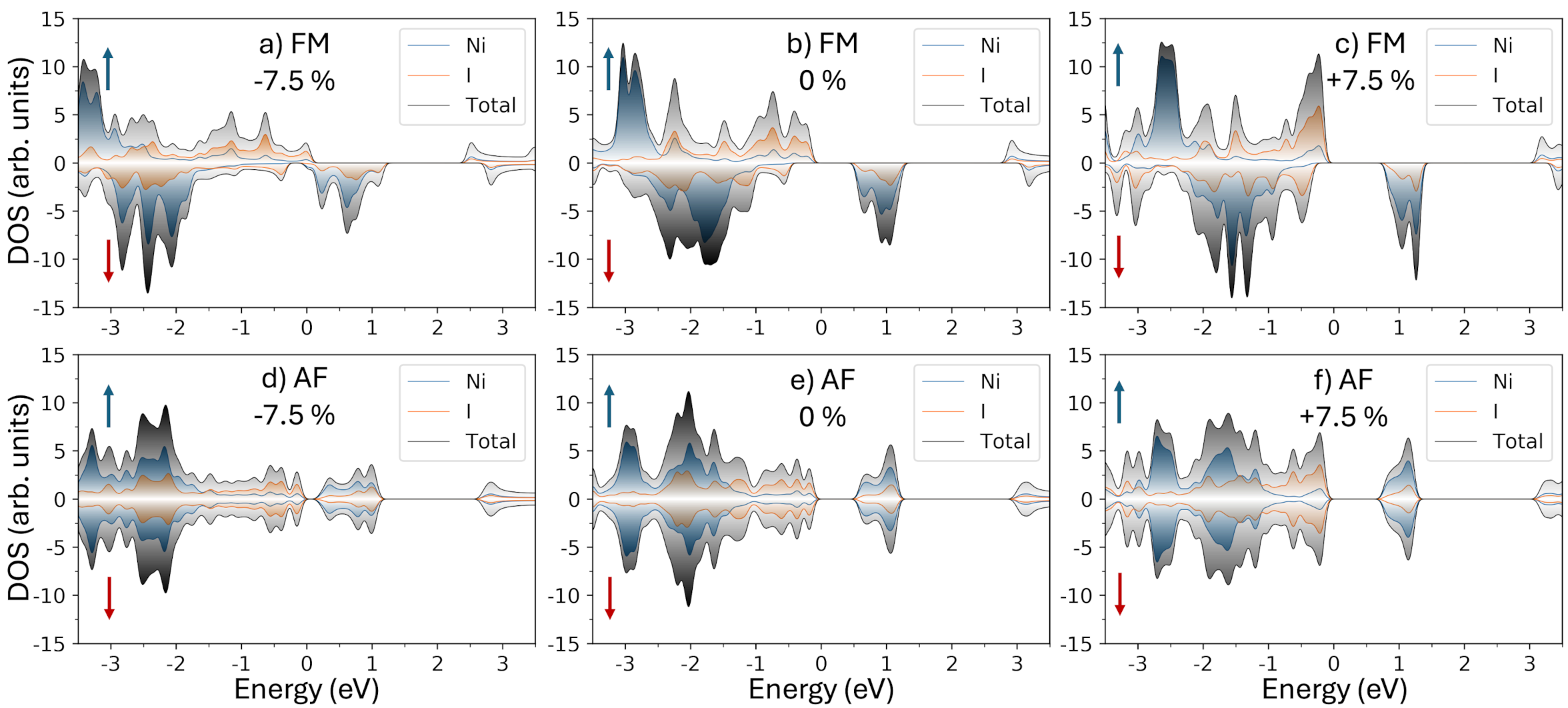} 
\caption{Spin-projected local density of states (DOS) for NiI$_2$ in both ferromagnetic (\textbf{top panel}) and antiferromagnetic (\textbf{bottom panel}) configurations under different strain conditions ($-7.5\%$, $0\%$, $+7.5\%$). The plots display the partial contributions from Ni (blue) and I (orange) atoms, alongside the total density of states (black), with the energy axis referenced to the Fermi level at 0 eV. Positive DOS values represent the spin-up projection, while negative values correspond to spin-down.}
\label{Fig:ldos}
\end{figure*}

\subsubsection*{Electronic relaxation time}

The electronic relaxation time $\tau$ was estimated using the semiclassical relation $\tau = \mu m^*/e$, where $\mu$ is the carrier mobility, $m^*$ is the effective mass expressed in units of the electron mass $m_e$, and $e$ is the elementary charge~\cite{neamen2003semiconductor,zhou2023enhanced, tomar2018thermoelectric}. The effective masses were calculated by fitting the band edges to parabolic functions derived from the first-principles band structure.
Since experimental measurements of mobility in monolayer NiI$_2$ are not yet available, we adopt a representative value of $\mu = 50\,\mathrm{cm}^2/\mathrm{V\cdot s}$, consistent with reported mobilities for related two-dimensional magnetic semiconductors. For instance, CrI$_3$ and similar systems typically exhibit mobilities ranging from $20$ to $80\,\mathrm{cm}^2/\mathrm{V\cdot s}$, depending on temperature, defects, and doping~\cite{neamen2003semiconductor, li2020recent, wu2019strain, shcherbakov2018raman}.
Using this estimate, we obtain $\tau = 22.7\,\mathrm{fs}$ for electrons and $10.8\,\mathrm{fs}$ for holes in the FM phase, corresponding to effective masses of $m^*_e = 0.7975\,m_e$ and $m^*_h = 0.3785\,m_e$. In the AF phase, the slightly larger effective masses ($m^*_e = 0.808\,m_e$, $m^*_h = 0.4455\,m_e$) yield relaxation times of $23.0\,\mathrm{fs}$ and $12.7\,\mathrm{fs}$, respectively. This modest increase aligns with the expected band flattening induced by antiferromagnetic interactions~\cite{han2020enhanced, tomar2018thermoelectric}.
Compared to other 2D magnets, these values are relatively short. For example, monolayer CrI$_3$ exhibits relaxation times in the range of $30$-$70\,\mathrm{fs}$~\cite{wu2019strain}. The shorter $\tau$ in NiI$_2$ may indicate stronger scattering, possibly linked to its complex magnetic landscape, including spin-spiral and helimagnetic order~\cite{amini2024atomic, nii2_multiferroic_osti}, which can enhance both electron-phonon and spin-mediated scattering processes~\cite{shcherbakov2018raman, han2020enhanced, canetta2024impact}.

In contrast, the Seebeck coefficients $S$ exhibit similar magnitudes in both magnetic phases, indicating that magnetic ordering has a relatively minor effect on the thermopower. This contrasts with its pronounced influence on charge transport, which will be evidenced by the substantial changes in $\sigma/\tau$ between FM and AF states~\cite{tomar2018thermoelectric}.

\subsection{Mechanical strain}

\begin{figure}[h!]
\centering
\includegraphics[clip,width=0.45\textwidth,angle=0]{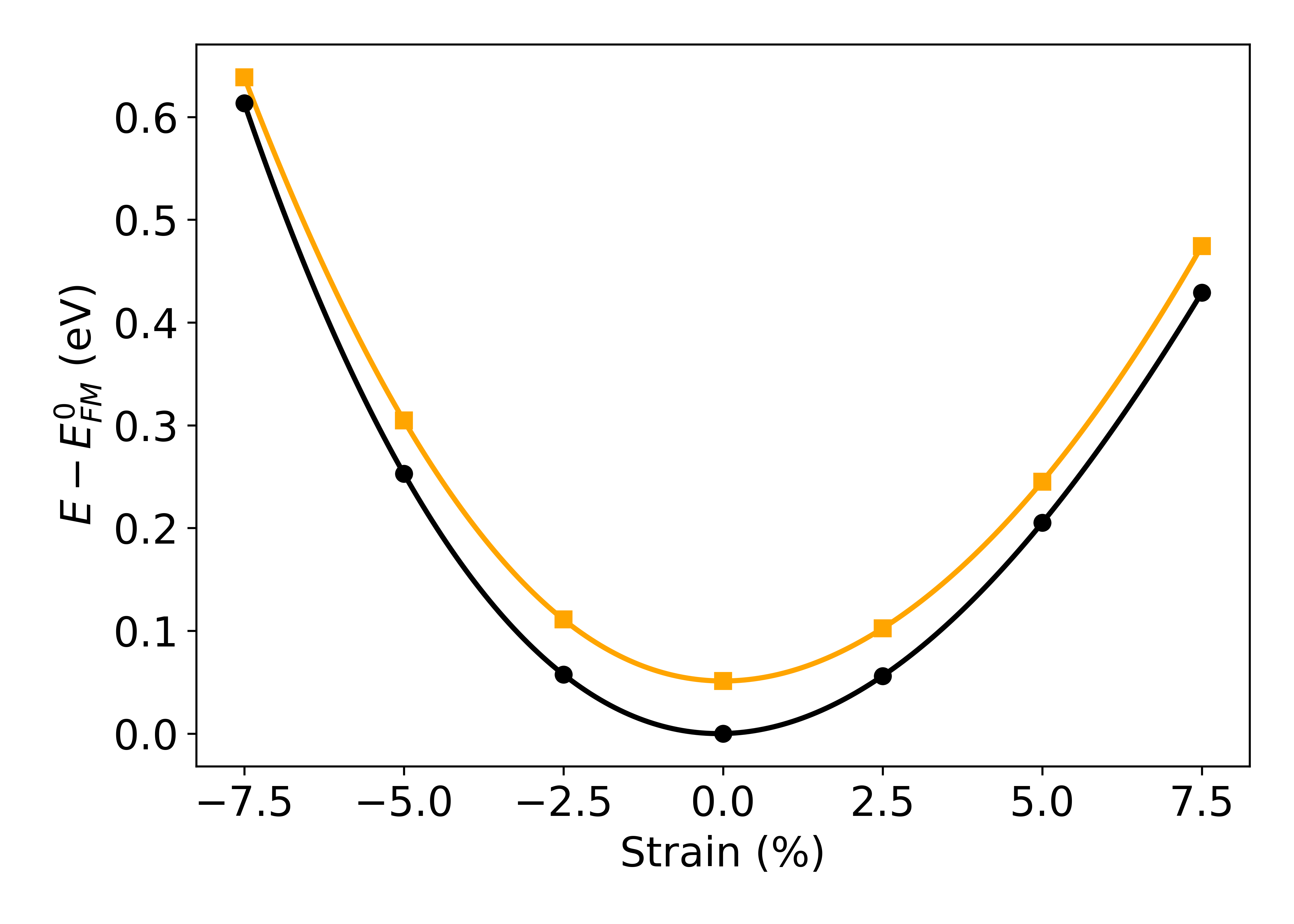} 
\includegraphics[clip,width=0.45\textwidth,angle=0]{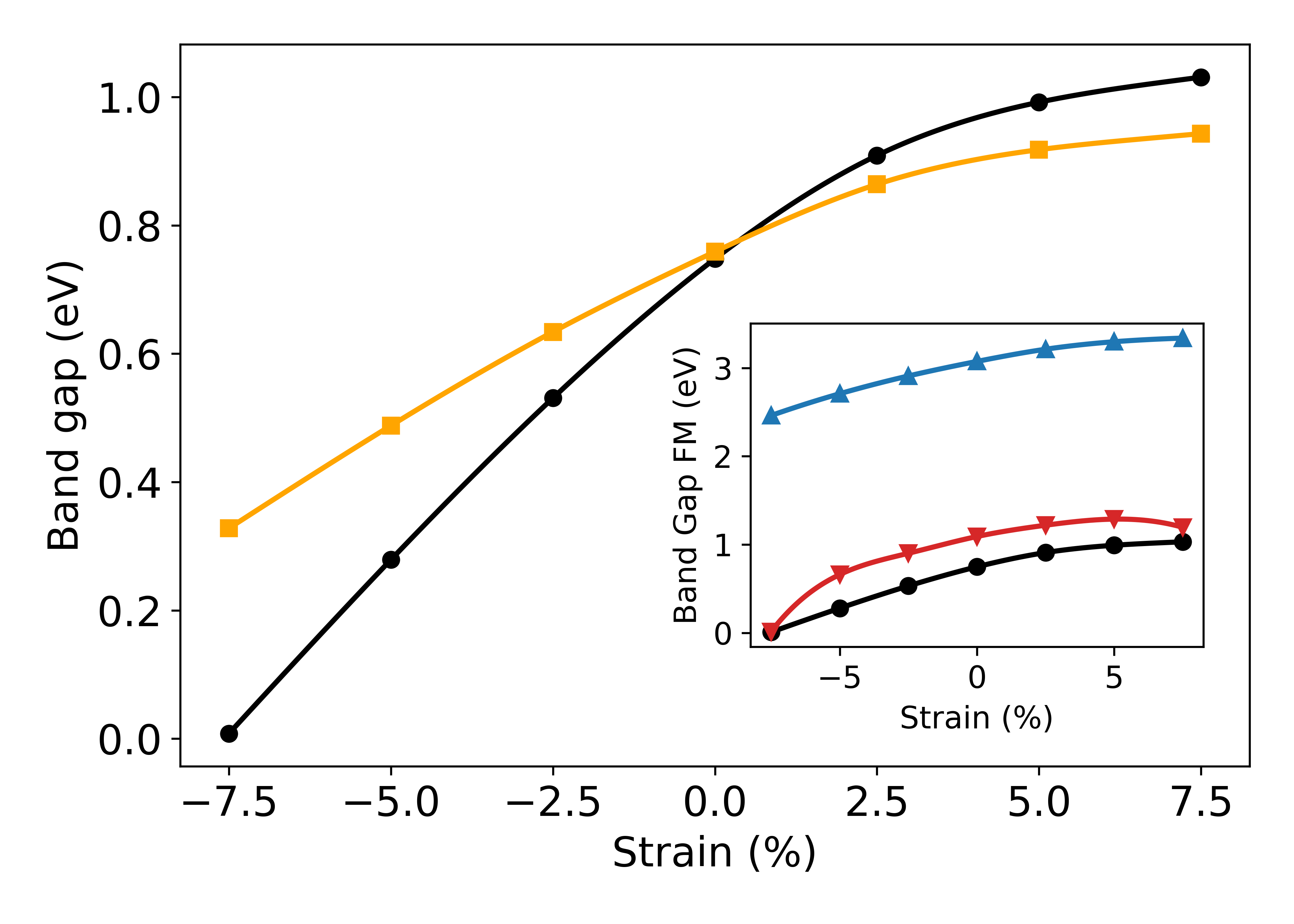} 
\caption{Relative energies of the ferromagnetic (black lines) and antiferromagnetic (orange lines) configurations of monolayer NiI\(_2\) as a function of biaxial strain (\textbf{top panel}), referenced to the ferromagnetic ground-state energy at 0\% strain. The \textbf{ bottom panel} shows the evolution of the electronic band gap for both magnetic phases under strain. The inset displays the spin-resolved band gap components for the FM configuration, with blue and red lines indicating the spin-up and spin-down channels, respectively.}
\label{Fig:strainFMAF}
\end{figure}

Figure~\ref{Fig:strainFMAF} shows the relative energies (top panel) of the FM and AF configurations of NiI$_2$ as a function of biaxial strain and the corresponding band gaps (bottom panel). We define biaxial strain as the percentage change in the in-plane lattice vectors relative to their equilibrium values. Thus, we use the definition of strain as $\varepsilon = (a -a_0)/a_o \times 100\%$, where $a_0$ and $a$ are the lattice parameters of the unstrained and strained structures, respectively. Positive and negative values of $\varepsilon$ correspond to tensile and compressive strain,  respectively, and range between $-7.5\%\leq \varepsilon \leq +7.5\%$. 

To investigate the influence of mechanical deformation on the electronic and thermoelectric properties of monolayer NiI$_2$, we applied biaxial strain in the range of $\pm 7.5\%$. To the best of our knowledge, no experimental studies have yet explored strain effects specifically in monolayer NiI$_2$. However, similar strain magnitudes have been widely considered in recent theoretical works focused on NiI$_2$ and other two-dimensional materials. Strain values in the range of $\pm 4\%$ to $\pm 8\%$ are typically used in first-principles studies to examine stability, magnetic ordering, and transport behavior under deformation~\cite{han2020enhanced,ni2025plane,ghojavand2024strain}. These levels are generally regarded as physically accessible in 2D systems through experimental techniques such as substrate-induced strain or flexible supports~\cite{dai2019strain}.
Therefore, the strain range adopted in this work is consistent with the existing theoretical literature, lies within a mechanically reasonable regime, and does not compromise the validity of the electronic and thermoelectric trends reported.

While epitaxial strain is known to induce structural phase transitions in perovskite oxides and other correlated systems~\cite{lu2016epitaxial,schlom2014elastic,lee2011coupled,leon2025strain}, monolayer NiI$_2$ is a van der Waals material with comparatively weak interlayer interactions and enhanced structural flexibility. Although surface reconstructions have been observed in ultrathin NiI$_2$ films grown on substrates~\cite{komarov2021new}, such reconstructions have not, to the best of our knowledge, been reported in isolated monolayers. Furthermore, no structural or dynamical instabilities have been predicted under moderate strain in recent first-principles studies, which consistently show that monolayer NiI$_2$ retains its symmetry and mechanical integrity across the relevant strain window~\cite{han2020enhanced,ni2025plane,ghojavand2024strain}.
It is important to note that the insulator-to-metal transition observed under compressive strain in our study does not involve any structural phase transition. Instead, it originates from enhanced hybridization between Ni $d$-orbitals and I $p$-orbitals, which leads to a spin-selective gap closure in the spin-down channel. As a result, the system exhibits a half-metallic ground state. This electronically driven transition differs fundamentally from the metallization observed in elemental semiconductors such as Si and Ge, where the gap closure is typically accompanied by a structural transformation under high pressure.

In the absence of mechanical strain, the ferromagnetic configuration (black curve) is energetically favored over the antiferromagnetic one, with its ground state lying 51.15\,meV per unit cell below that of the AF configuration (orange curve). While the in-plane interatomic distances remain nearly identical between both magnetic orders, the AF configuration exhibits shorter out-of-plane I-I distances, which contributes to its higher total energy relative to the FM configuration. Regarding the behavior of the total band gap in both magnetic orders, in the absence of strain, both systems share the same indirect electronic band gap, which undergoes distinct evolutions as mechanical strain is applied.

Under the application of biaxial strain, the ferromagnetic (FM) configuration consistently remains the energetically preferred magnetic state across the entire range of strain values investigated. However, the relative stability of the FM state with respect to the antiferromagnetic (AF) configuration exhibits a clear dependence on the strain magnitude. Specifically, under compressive strain, the energy difference between the FM and AF states progressively decreases, indicating a gradual weakening of the FM state's dominance. For example, at a compressive strain of $-7.5\%$, the energy difference is reduced to approximately 25 meV per unit cell. Despite this reduction, the FM phase still constitutes the ground state, suggesting that while strain influences magnetic stability, it does not induce a magnetic phase transition within the explored strain range.

The bottom panel of Figure~\ref{Fig:strainFMAF} presents the evolution of the electronic band gaps for both FM and AF configurations under biaxial strain. In both magnetic phases, tensile strain leads to an increase in the band gap, consistent with enhanced electronic localization due to reduced orbital overlap. Despite this common trend, the AF configuration consistently exhibits slightly narrower gaps compared to the FM phase. Under compressive strain, however, the FM band gap decreases more rapidly and nonuniformly, ending in a semiconductor-to-metal transition at $\epsilon = -7.5\%$.

The spin-resolved band gaps reported in Fig.~\ref{Fig:strainFMAF} are calculated within each spin channel as the energy difference between the valence band maximum and conduction band minimum, while the total gap corresponds to the smallest energy difference between the highest occupied and lowest unoccupied states, irrespective of spin. A more detailed analysis, shown in the inset of Fig.~\ref{Fig:strainFMAF} and corroborated by the spin-resolved local density of states (LDOS) in Fig.~\ref{Fig:ldos}, reveals that the metallic transition is driven by the closure of the spin-down gap, while the spin-up channel remains gapped. The spin-down conduction band minimum, initially located near 1~eV in the unstrained case, progressively shifts downward and broadens under increasing compression, particularly near the $M$ point. This state, primarily composed of hybridized Ni-$d$ and I-$p$ orbitals, is highly dispersive and exhibits enhanced sensitivity to strain-induced modifications in orbital overlap. In contrast, the spin-up conduction states are more localized and less dispersive, maintaining a finite gap throughout the strain range. These observations confirm the emergence of a half-metallic phase in the FM configuration, characterized by metallic conduction in one spin channel and an insulating gap in the other.

Figure~\ref{Fig:ldos} further shows the contrasting LDOS response of both magnetic configurations under $\pm7.5\%$ strain. In the AF phase, spin degeneracy is preserved across the entire strain range, and the modulation of the band gap occurs symmetrically, with comparable contributions from both atomic species. This behavior reflects the robustness of the AF magnetic ordering against spin-dependent distortions. In contrast, the FM phase exhibits a pronounced spin asymmetry under strain. Under tensile deformation, the band gap increases while preserving spin polarization, and the states near 1~eV become more localized, particularly on Ni atoms. Under compressive strain, however, the LDOS reveals a semiconductor-to-metal transition, dominated by the broadening and downward shift of Ni spin-down and I spin-up states, which intersect the Fermi level; this shows the central role of Ni-$d$/I-$p$ hybridization in mediating the transition.

The strain-induced asymmetry and selective gap closure of one spin channel distinguishes NiI$_2$ from other two-dimensional magnetic semiconductors such as CrI$_3$. While CrI$_3$ undergoes strain-induced magnetic phase transitions due to interlayer coupling, its electronic structure remains spin-symmetric and insulating across a broad range of strains~\cite{leon2020strain}. In contrast, monolayer NiI$_2$ displays a strain-driven electronic transition into a robust half-metallic phase, demonstrating its potential as a mechanically tunable platform for spin-filtering and other spintronic functionalities.

\section{Thermoelectric properties}

\begin{figure}[]
\centering
\includegraphics[clip,width=0.45\textwidth,angle=0]{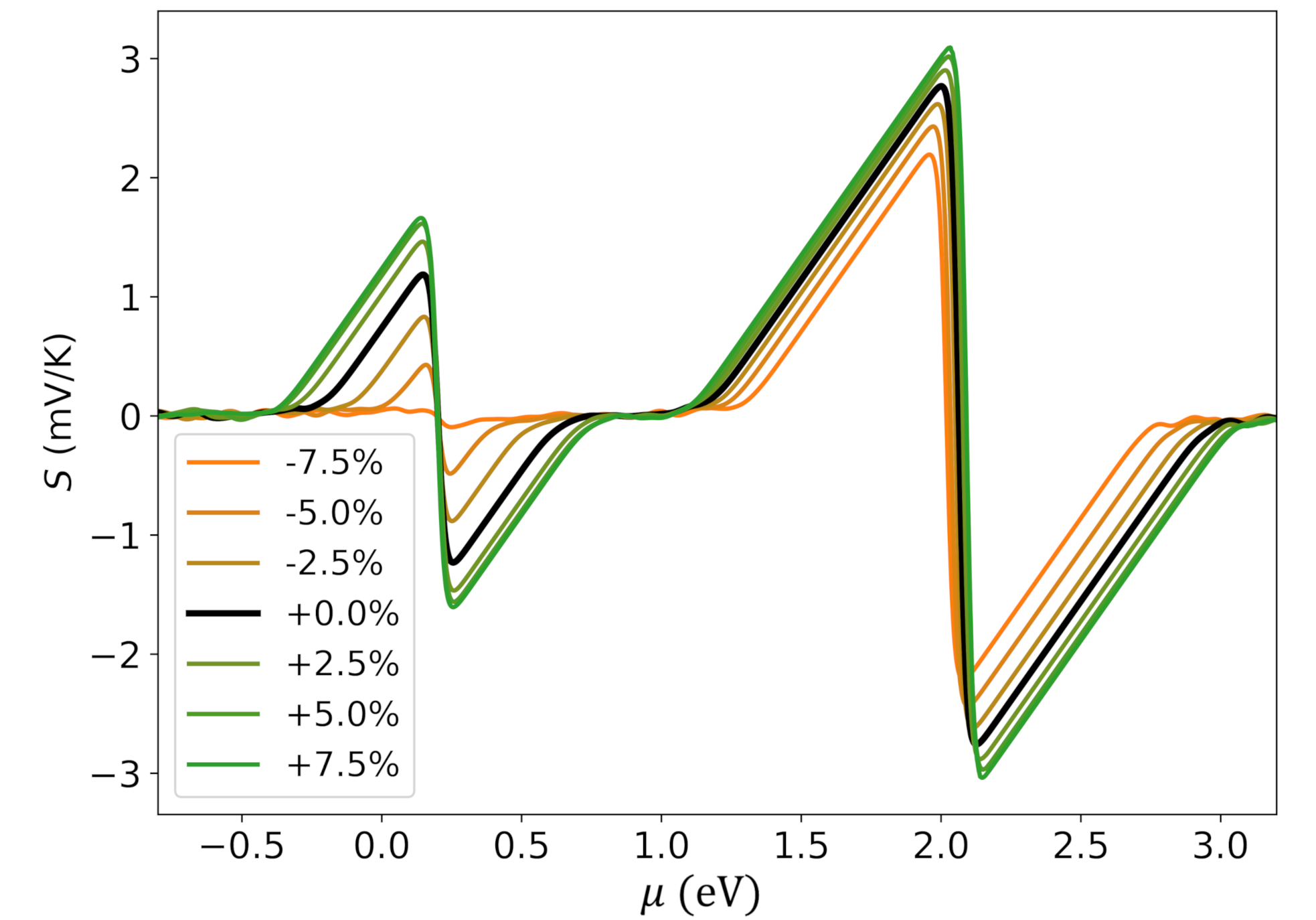} 
\includegraphics[clip,width=0.45\textwidth,angle=0]{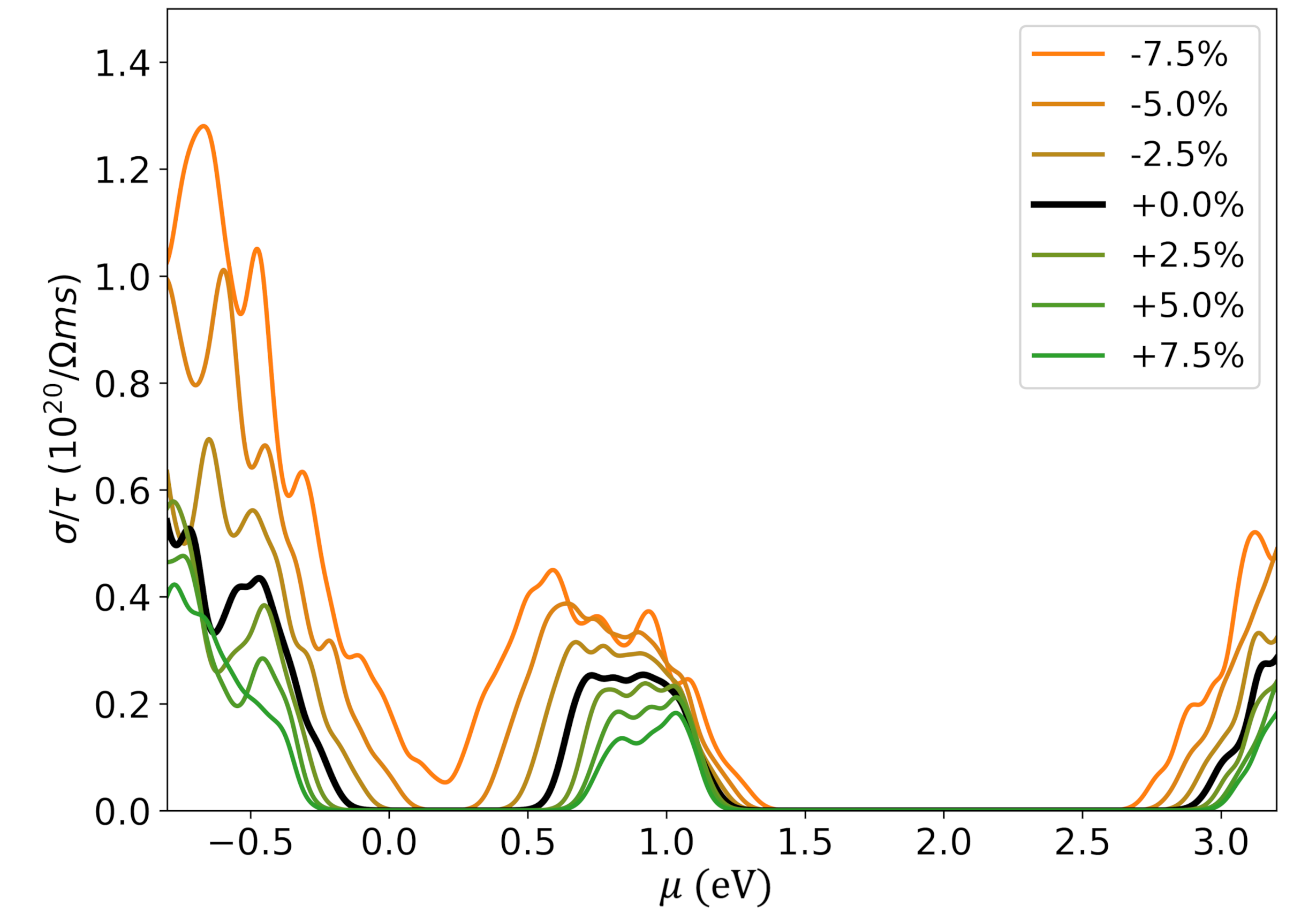} 
\caption{Seebeck coefficient \( S \) (\textbf{top panel}) and electrical conductivity \( \sigma/\tau \) (\textbf{bottom panel}) as functions of energy relative to the Fermi level, \( E - E_F \), for the ferromagnetic configuration of monolayer NiI\(_2\) under various biaxial strain values ranging from \(-7.5\%\) to \(+7.5\%\), calculated at a fixed temperature of $300\,$K. }
\label{Fig:S}
\end{figure}

We investigate the effect of mechanical strain on the thermoelectric properties of NiI$_2$, focusing on how strain influences its transport behavior, particularly in the ferromagnetic phase. Figure~\ref{Fig:S} presents the Seebeck coefficient $S$ (top panel) and the electrical conductivity normalized by relaxation time, $\sigma/\tau$ (bottom panel), for the ferromagnetic NiI$_2$ monolayer. These properties are shown as functions of energy relative to the Fermi level, $E - E_F$, under various biaxial strains ranging from $-7.5\%$ to $+7.5\%$, at a temperature of $300\,$K.

Under tensile strain ($\varepsilon > 0\%$), the Seebeck coefficient $S$ preserves its typical semiconducting behavior, characterized by prominent peaks near the edges of the conduction and valence bands. This behavior reflects the asymmetric energy-dependent transport of charge carriers near the band extrema. As compressive strain is gradually introduced ($\varepsilon < 0\%$), the overall magnitude of $S$  near the Fermi level decreases, indicative of a narrowing band gap and corresponding modifications in the electronic band structure, including changes to the carrier effective masses. Notably, at a compressive strain of $-7.5\%$, the ferromagnetic NiI$_2$ monolayer undergoes a semiconductor-to-metal transition. This transition eliminates the energy filtering effect typically responsible for thermoelectric voltage generation, resulting in a vanishing Seebeck coefficient—consistent with the expected behavior of a conventional metallic system.

In the case of electrical conductivity, $\sigma/\tau$ (shown in the bottom panel), the effect of mechanical strain is more pronounced under compression compared to the tensile regime. Although tensile strain leads to a reduction in conductivity—primarily due to the increased distance between nickel and iodine atoms—the overall behavior remains similar to that of the unstrained system, with only slight decreases in conductivity observed for $\varepsilon > 0$.
In contrast, when the system is progressively compressed, two distinct behaviors emerge. First, the conductivity increases with greater compressive strain, mainly as a result of the reduced Ni-Ni and I-I atomic distances, which enhance the overlap between the transition metal atomic orbitals. Second, the energy gap begins to close around the Fermi level, leading to a semiconductor-to-metal transition at $\varepsilon = -7.5\%$, driven by the overlap between the atomic orbitals of the transition metal and those of the ligands, as previously discussed.
This behavior is consistent across temperatures across temperatures. This suggests that the effects of strain are stable and not strongly influenced by temperature changes, making the results useful for thermoelectric applications.

\begin{figure}[]
\centering
\includegraphics[clip,width=0.45\textwidth,angle=0]{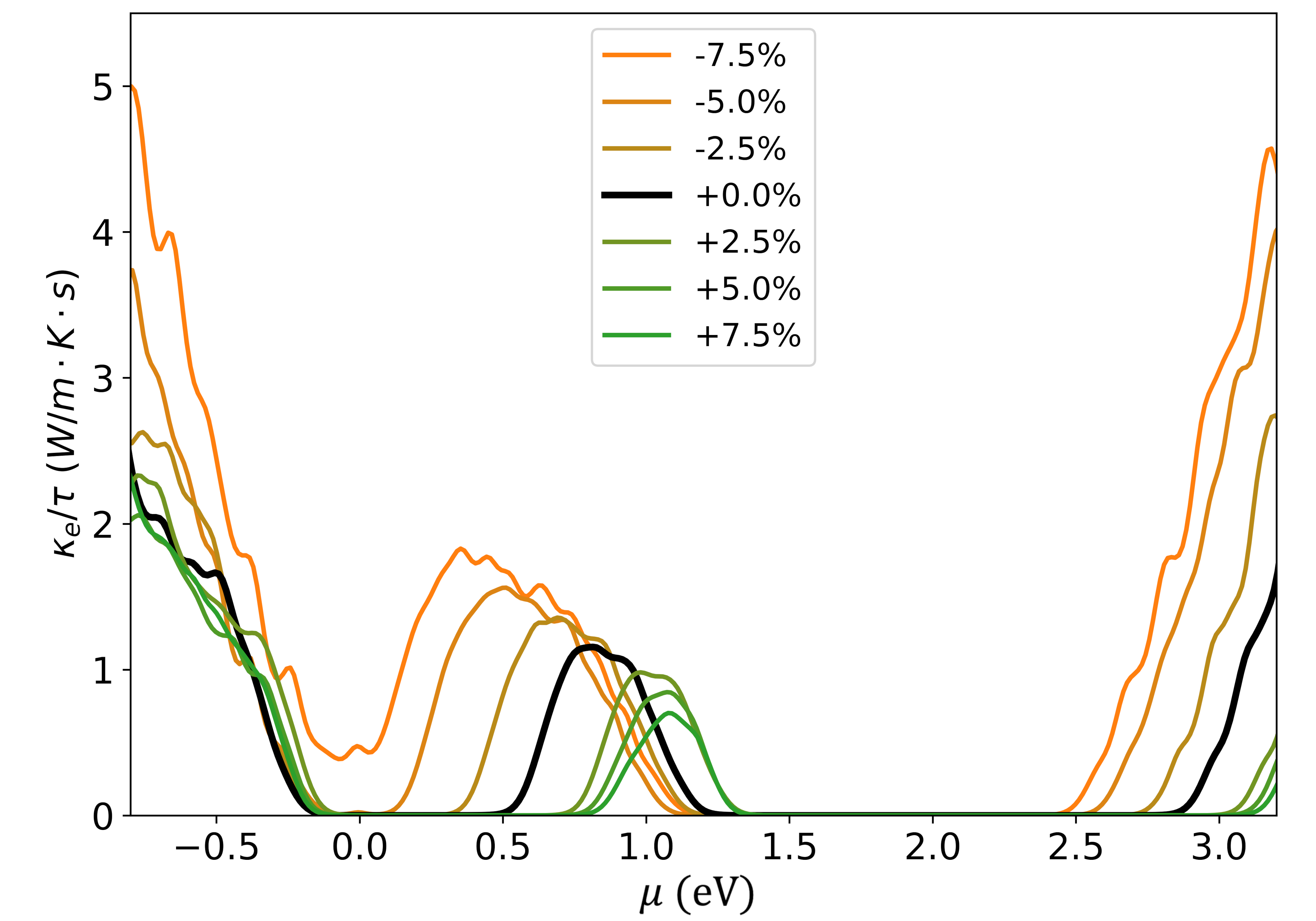} 
\includegraphics[clip,width=0.46\textwidth,angle=0]{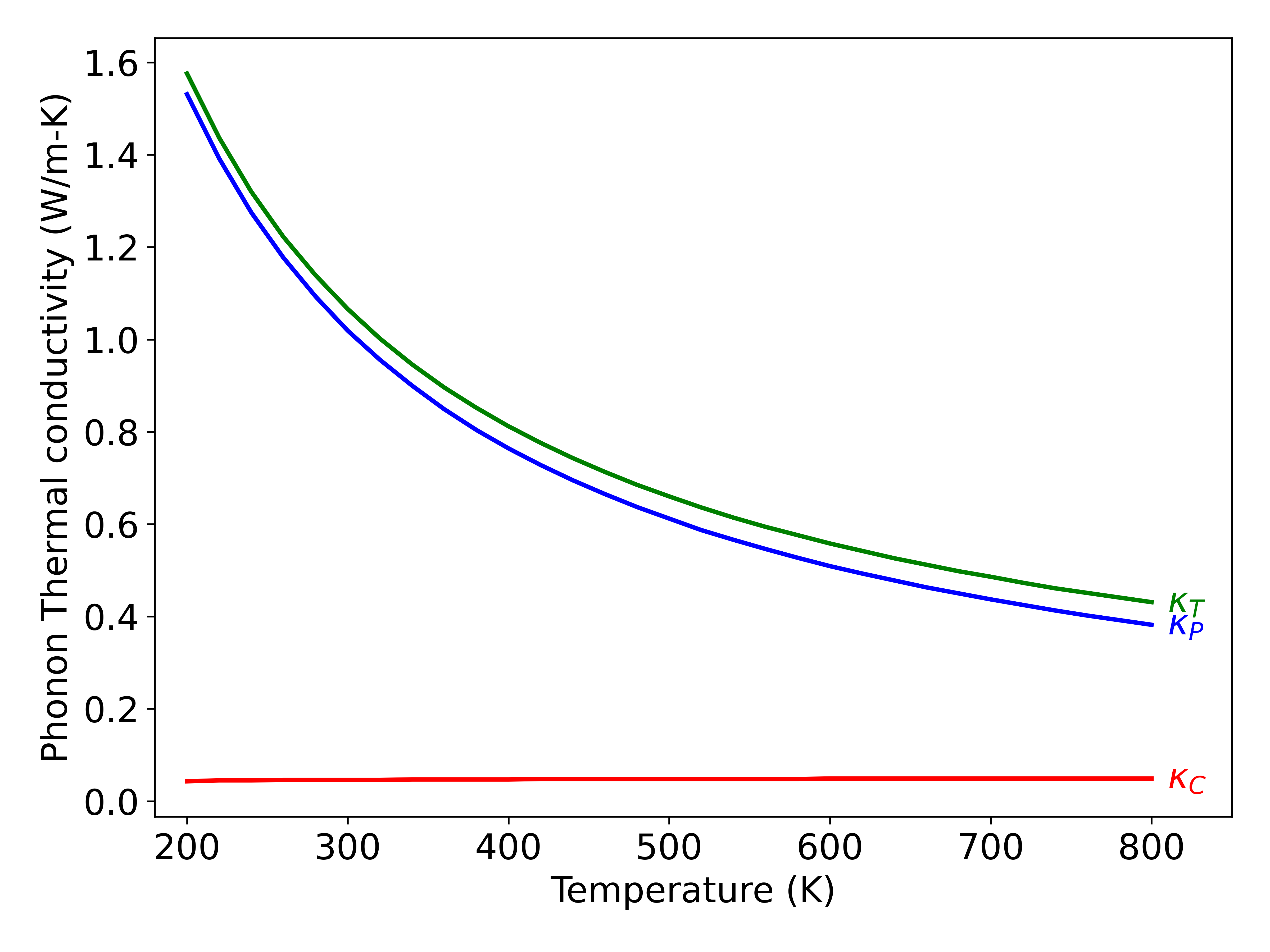} 
\caption{Electronic thermal conductivity \( \kappa_{e}/\tau \) (\textbf{top panel}) as a function of chemical potential for the ferromagnetic configuration of monolayer NiI\(_2\) at $300\,$K. Phonon thermal conductivity at \( \kappa_{\mathrm{ph}} \) (\textbf{bottom panel}) at $\varepsilon = 0$, as a function of temperature. According to the Wigner transport equation, \( \kappa_{\mathrm{ph}} \) can be decomposed into a particle-like contribution \( \kappa_{P} \) and a coherence contribution \( \kappa_{C} \), which accounts for wave-like tunneling of phonons between bands with energy separations smaller than their linewidths.}
\label{Fig:kappa}
\end{figure}

Regarding the thermal conductivity of NiI$_2$, we have calculated both the electronic contribution $\kappa_e$ and the phonon contribution $\kappa_{\mathrm{ph}}$ as functions of biaxial strain and temperature. The top panel of Figure~\ref{Fig:kappa} shows the behavior of $\kappa_e$ under varying strain at $300\,$K. Under tensile strain, $\kappa_e$ remains relatively unchanged compared to the unstrained case, with only a minor shift in the spectral features, attributed to the increased localization of spin-down bands, particularly pronounced at $\varepsilon = +7.5\%$. In contrast, under compressive strain, the system exhibits a substantial increase in $\kappa_e$, especially near the Fermi level. This enhancement correlates with the strain-driven semiconductor-to-metal transition and reflects an overall rise in electronic transport activity.
The differences between the evolution of $\kappa_e$ and the corresponding electrical conductivity $\sigma$ under strain motivates a deeper examination of the validity of the Wiedemann-Franz law in this system\cite{chester1961law}.

The bottom panel of Figure~\ref{Fig:kappa} shows the phonon thermal conductivity, $\kappa_{\mathrm{ph}}$, as a function of temperature. As temperature increases, $\kappa_{\mathrm{ph}}$ exhibits a gradual decrease, a behavior commonly observed in crystalline materials, including two-dimensional systems \cite{Gandi_2016}. This trend can be attributed to enhanced phonon-phonon scattering at elevated temperatures. Within the framework of the Wigner transport equation, $\kappa_{\mathrm{ph}}$ can be decomposed into a particle-like contribution, $\kappa_{P}$, and a coherence contribution, $\kappa_{C}$. The latter accounts for wave-like tunneling processes of phonons between closely spaced bands, thereby reflecting the quantum nature of heat transport in low-dimensional materials. Furthermore, in general terms, small mechanical strain will not significantly affect the phonon contribution to the thermal conductance, since the typical wavelength of an acoustic phonon (the dominant terms) is much larger than the interatomic spacing of the material.

\begin{figure}[htbp!]
\centering
\includegraphics[clip,width=0.45\textwidth,angle=0]{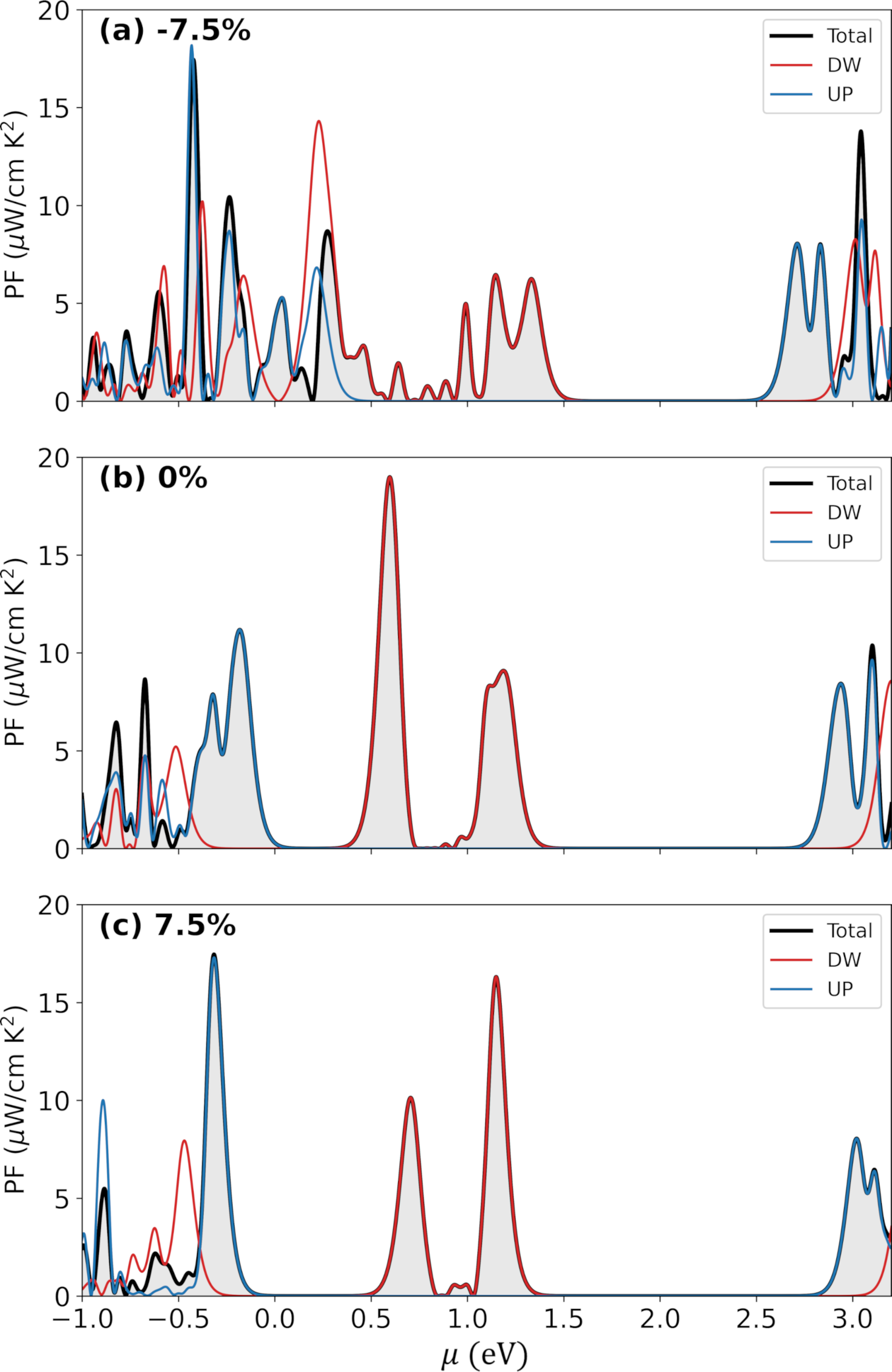} 
\caption{Power factor for the ferromagnetic configuration of monolayer NiI\(_2\) under different biaxial strain values: (a) \(-7.5\%\), (b) \(0\%\), and (c) \(+7.5\%\). The black curve (with a shaded region) indicates the total PF, while the red and blue curves correspond to the spin-up and spin-down contributions, respectively.}
\label{Fig:PF}
\end{figure}

Figure~\ref{Fig:PF} presents the power factor for the FM NiI$_2$ monolayer under various strain conditions: (a) $-7.5\%$, (b) $0\%$, and (c) $+7.5\%$. The power factor, defined in Eq. \ref{eq:PF}, combines the Seebeck coefficient ($S$) and the electrical conductivity ($\sigma$), providing a key metric for thermoelectric performance. It is particularly valuable for analyzing spin-dependent thermoelectric properties, as it explicitly captures how differences in spin-resolved transport channels affect the material's overall efficiency.

In the absence of strain [panel (b)], the PF shows distinct peaks in both spin channels, with the spin-down contribution dominating near the conduction band edge (around $\approx 1$ eV). In contrast, the spin-up  channel is more prominent below the Fermi level. When tensile strain ($+7.5\%$) is applied [panel (c)], the PF increases significantly, especially in the spin-up channel, reflecting enhanced thermoelectric performance associated with band-gap widening and improved carrier mobility. In contrast, under compressive strain ($-7.5\%$) [panel (a)], the PF peaks shift closer to the Fermi level, and multiple sharp peaks emerge due to increased carrier density and reduced band gap, consistent with the previously observed metallic transition. Under these conditions, the PF shows notable differences between spin channels, indicating strong spin polarization and suggesting that strain engineering can effectively tune spin-dependent thermoelectric properties in NiI$_2$.

\subsubsection*{Anomalous breakdown of the Wiedemann-Franz law}

To further assess the strain dependence of electronic thermal transport, we computed the Lorenz number defined in Eq.~\ref{eq:L}, which quantifies the ratio between heat and charge conduction. Table~\ref{tab:lorenz} shows the average Lorenz number $\langle L \rangle$ computed within a narrow energy window around the Fermi level ($|E - E_F| < 0.1$\,eV), a commonly used range in thermoelectric studies~\cite{usui2017enhanced} that captures the dominant low-energy transport processes in narrow-gap systems.

At zero strain, $\langle L \rangle$ is close to the Sommerfeld value $L_0$, consistent with conventional semiconducting behavior. Under tensile strain, $\langle L \rangle$ steadily decreases to $0.36\,L_0$ at $\varepsilon = +7.5\%$, indicating enhanced carrier localization and reduced thermal conductivity relative to electrical conductivity. Such suppression reflects a breakdown of the free-electron assumptions underpinning the Wiedemann-Franz law, a phenomenon also observed in 2D semiconductors such as FePS$_3$ and CrI$_3$, where $L$ typically remains within $1.1\,L_0$~\cite{guo2025spin}.

In contrast, compressive strain induces a sharp, non-monotonic enhancement of $\langle L \rangle$, peaking at $7.17\,L_0$ for $\varepsilon = -2.5\%$. Interestingly, this maximum does not coincide with the most metallic state at $\varepsilon = -7.5\%$, but instead emerges at an intermediate strain where the band structure is undergoing significant reconstruction. This behavior suggests that the strongest violations of the Wiedemann-Franz law arise not in fully metallic regimes but near electronic crossover regions, possibly linked to a strain-induced Lifshitz transition, where the Fermi surface topology changes and spin-selective band crossings alter the relative contributions of thermal and electrical channels. Such conditions can introduce resonant states or energy filtering effects that disproportionately enhance $\kappa_e$ while leaving $\sigma$ relatively unchanged. These features align with theoretical predictions of strong deviations near electronic criticality~\cite{kubala2008violation,tu2023wiedemann}.

Experimental evidence corroborates the plausibility of such anomalies. In the heavy-fermion metal CeCoIn$_5$, a pronounced violation of the Wiedemann-Franz law was observed below the superconducting critical field, with $L/L_0$ exceeding 4.2 at low temperatures~\cite{jaoui2018departure}. This behavior was attributed to the decoupling of electronic heat and charge transport channels induced by quantum fluctuations. Similarly, in monolayer WTe$_2$, gating-induced Lifshitz transitions were shown to suppress the Lorenz number to as low as $0.25\,L_0$~\cite{wang2025unusual}, facilitating unconventional thermoelectric performance. In both cases, the anomalous Lorenz number reflects a significant restructuring of the Fermi surface and the emergence of selective scattering pathways, mechanisms that closely parallel the spin-resolved band-crossing and filtering effects identified in strained NiI$_2$.

Deviations from the Wiedemann-Franz law have also been reported in systems with "pudding-mold'' band structures~\cite{usui2017enhanced}, where coexisting flat and dispersive regions yield strong energy dependence in transport coefficients. A comparable mechanism may be at play in strained NiI$_2$, where spin-resolved asymmetries and band reconstruction disrupt the conventional link between $\kappa_e$ and $\sigma$.

\begin{table}[h!]
\centering
\caption{Strain-dependent average Lorenz number $\langle L \rangle$ evaluated within a $\pm 0.1$\,eV window around the Fermi level at $300\,$K for the ferromagnetic configuration of monolayer NiI$_2$. Values are shown in absolute units (W$\cdot\Omega$/K$^2$) and relative to the Sommerfeld limit $L_0 = 2.44 \times 10^{-8}$\,W$\cdot\Omega$/K$^2$.\\}
\label{tab:lorenz}
\begin{tabular}{c|c|c}
\hline
Strain (\%) & $\langle L \rangle$ (W$\cdot\Omega$/K$^2$) & $\langle L \rangle / L_0$ \\
\hline
$-7.5$ & $3.35 \times 10^{-8}$ & $1.37$ \\
$-5.0$ & $1.15 \times 10^{-7}$ & $4.71$ \\
$-2.5$ & $1.75 \times 10^{-7}$ & $7.17$ \\
$+0.0$ & $1.80 \times 10^{-8}$ & $0.74$ \\
$+2.5$ & $1.56 \times 10^{-8}$ & $0.64$ \\
$+5.0$ & $1.09 \times 10^{-8}$ & $0.45$ \\
$+7.5$ & $8.80 \times 10^{-9}$ & $0.36$ \\
\hline
\end{tabular}
\end{table}

From a theoretical perspective, such violations imply that charge and heat currents originate from different parts of the electronic structure, incompatible with Fermi-liquid assumptions. Analogous effects occur in quantum dot systems~\cite{kubala2008violation}, where Coulomb blockade yields highly energy-selective transport, and in graphene near the Dirac point~\cite{tu2023wiedemann}, where bipolar conduction enhances thermal currents independently of charge flow.

Beyond fundamental interest, these anomalies have direct technological implications. The Wiedemann-Franz law enforces a trade-off in conventional thermoelectric materials: improving electrical conductivity ($\sigma$) inevitably increases electronic thermal conductivity ($\kappa_e$), thus limiting the figure of merit $ZT = \sigma S^2 T / (\kappa_e + \kappa_{ph})$. Materials where $L \ll L_0$ can break this constraint, allowing for efficient thermoelectric performance without the thermal penalty. In this scenario, monolayer NiI$_2$ is particularly promising: it not only violates the Wiedemann-Franz law but does so reversibly under strain, enabling precise tuning of transport properties. 
Importantly, strain levels on the order of a few percent are routinely achieved in 2D materials using flexible substrates, epitaxial mismatch, or local AFM indentation~\cite{peng2020strain,yu2024manipulating}, supporting the experimental feasibility of our predictions. NiI$_2$ thus emerges as a compelling platform for strain-tunable thermoelectric and spin-caloritronic technologies, where independent control over heat and charge flow is essential.

\section{Final Remarks}

We have demonstrated that biaxial strain fundamentally reshapes the quantum transport landscape of monolayer NiI$_2$. A moderate compressive strain of $-7.5\%$ drives a semiconductor-to-half-metal transition, accompanied by a non-monotonic, sevenfold enhancement of the Lorenz number, revealing a dramatic violation of the Wiedemann-Franz law. This anomaly signals a departure from conventional Fermi-liquid behavior and originates from the strain-sensitive hybridization of Ni-$d$ and I-$p$ orbitals, which selectively close the spin-down band gap while maintaining spin-up semiconducting character.

This spin-resolved asymmetry leads to a half-metallic state, where conduction and valence states are dominated by opposite spin channels, allowing for the selective manipulation of spin-polarized transport. Crucially, the decoupling between charge ($\sigma$) and heat ($\kappa_e$) transport under strain opens new avenues for overcoming thermoelectric trade-offs, where enhancing electrical conductivity does not incur the usual thermal penalty.

Altogether, these findings establish monolayer NiI$_2$ as a multifunctional 2D platform where spin, charge, and heat transport can be decoupled and modulated via mechanical strain. The observed departure from Fermi-liquid behavior, manifested in a giant, reversible violation of the Wiedemann-Franz law, introduces new design principles for thermoelectric and spin-caloritronic systems. In particular, the ability to independently tune electrical conductivity and thermal response opens pathways for mechanically gated thermoelectric switches and reconfigurable thermal logic devices. These results position NiI$_2$ as a model system for integrating quantum transport phenomena into functional device architectures.

\section*{Acknowledgments}
JWG and LR acknowledge the financial support from ANID-FONDECYT 1220700 (Chile).
JWG acknowledges financial support from ANID-FONDECY grant N. 1221301 (Chile).
Powered@NLHPC: This research was partially supported by the supercomputing infrastructure of the NLHPC (CCSS210001).

\bibliography{rsc} 
\bibliographystyle{rsc} 

\end{document}